\documentclass[aps,twocolumn,prl,showpacs,floatfix]{revtex4}

\usepackage{amsmath,fixmath}
\usepackage{graphicx}
\usepackage{amssymb}
\usepackage{bm}
\newcommand{\dd}{\mathrm{d}}
\newcommand{\ee}{\mathrm{e}}

\begin{document}

\title{A cluster Monte Carlo algorithm with a conserved order parameter}

\author{V.~Martin-Mayor} \affiliation{Departamento de F\'\i{}sica
  Te\'orica I, Universidad Complutense, 28040 Madrid, Spain} 
\affiliation{Instituto de Biocomputaci\'on y
  F\'{\i}sica de Sistemas Complejos (BIFI), Zaragoza, Spain.}

\author{D.~Yllanes}  \affiliation{Departamento de F\'\i{}sica
  Te\'orica I, Universidad Complutense, 28040 Madrid, Spain}
  \affiliation{Instituto de Biocomputaci\'on y
  F\'{\i}sica de Sistemas Complejos (BIFI), Zaragoza, Spain.}

\date{\today}

\begin{abstract}
We propose a cluster simulation algorithm for 
statistical ensembles with fixed order parameter. 
We use the tethered ensemble, which features Helmholtz's effective
potential rather than Gibbs's free energy, and in which
canonical averages are recovered with arbitrary accuracy. 
For the $D=2,3$ Ising
model  our method's critical slowing down 
is comparable to that of canonical cluster algorithms.
Yet, we can do more 
than merely reproduce canonical values. As an example,
we obtain a competitive value for the $3D$ 
Ising anomalous dimension from the maxima of the effective potential. 
\end{abstract}
\maketitle 
Monte Carlo simulations~\cite{MC} constitute
one of the most important modern tools
of theoretical physics.
In situations that defy an analytical treatment, a 
Monte Carlo computation succeeds by wandering 
randomly across the system's configuration space. 
Over the years many methods have been proposed in order to optimize
this sampling process. Here we combine
two: fixing some global parameter in order 
to guide the exploration of phase space and the use
of cluster update algorithms.

Cluster methods appear in the 1980s~\cite{CLUSTER-METHODS} as
an answer to the problem of critical slowing down~\cite{ZINN-JUSTIN}, 
the drastic deceleration of the dynamics
in the neighborhood of the critical point.
For a system of linear size $L$, the
characteristic times are $\tau \propto L^z$  (typically $z\approx 2$).
Only in very few situations
can one define an efficient cluster method, capable of achieving $z<1$.

Beyond the canonical ensemble setting, considering
global conservation laws is often useful 
(as in micromagnetic~\cite{KAWASAKI}
or microcanonical~\cite{MICROCANONICO} ensembles).
It is known that 
$z=4-\eta$  for locally conserved order parameter dynamics in Ising
models~\cite{HOHENBERG-HALPERIN} ($\eta$ is the anomalous
dimension).
For non-local conservation laws
$z$ is smaller~\cite{BRAY-TAMAYO-KLEIN,ANNETT-BANAVAR}.
Despite continued research on cluster methods~\cite{CLUSTER-MODERNOS},
the development of an efficient cluster method in this situation
has long been considered somewhat of a challenge~\cite{ANNETT-BANAVAR}. 

Here we present a working cluster algorithm 
with a globally conserved order parameter. 
We employ the Tethered Monte Carlo (TMC) framework~\cite{TMC},
which we briefly review.  TMC is a general
approach to reconstruct the effective
potential~\cite{VICTORAMIT,ZINN-JUSTIN}.
We demonstrate our cluster method in the standard benchmark of the 
$D=2,3$ Ising model. 
In the first case it outperforms the Metropolis 
version of~\cite{TMC},
while in the second it exhibits a dynamic critical
exponent compatible with that of the canonical Swendsen-Wang algorithm.

The tethered ensemble is similar to the micromagnetic one,
but instead of fixing the magnetization we couple it to 
a Gaussian `magnetostat' in order to define a new 
parameter, $\hat m$.  This ensemble arises 
from the canonical one through a Legendre transformation
that replaces the magnetic field $h$ by $\hat m$. 
Thus, Helmholtz's effective potential takes
the place of Gibbs's free energy. The 
main observable is the tethered magnetic field $\hat h$, 
considered as a function of $\hat m$.

We shall work on the Ising model in a cubic lattice 
of size $N=L^D$ and periodic boundary conditions, with partition function
($\langle\cdot,\cdot\rangle$: nearest neighbors)
\begin{equation}\label{eq:Z}
Z = \sum_{\{\sigma_{\boldsymbol x}\}} \exp\biggl[\beta \sum_{\langle \boldsymbol x, \boldsymbol y\rangle}
\sigma_{\boldsymbol x} \sigma_{\boldsymbol y}\biggr], 
\quad \sigma_{\boldsymbol x} = \pm  1.
\end{equation}
We shall consider its energy and magnetization,
\begin{align}
E&= Ne= -\frac1D \sum_{\langle \boldsymbol x,\boldsymbol y\rangle} \sigma_{\boldsymbol x}\sigma_{\boldsymbol y},&
M&= Nm = \sum_{\boldsymbol x} \sigma_{\boldsymbol x}.
\end{align}
We use lowercase for densities so that, for instance, $e$ is the 
energy per bond. Canonical averages are denoted by $\langle\cdot\rangle_\beta$,
as in the specific heat and susceptibility:
\begin{align}
C &= N \left[ \langle e^2\rangle_\beta - \langle e\rangle^2_\beta\right], &
\chi &= N \left[ \langle m^2\rangle_\beta - \langle m\rangle^2_\beta\right].
\end{align}

Note that the probability density (pdf) $p_1(m)$ 
is the sum of $N+1$ Dirac deltas. We smooth it 
by coupling $m$ to $N$ Gaussian demons
to build the tethered ensemble
\begin{equation}\label{eq:hatm}
\hat M = N \hat m =   M + \frac1{2} \sum_i \phi_i^2.
\end{equation}
As the $\phi_i$ are independent, $\hat m \simeq m+1/2$.
The definitions of the pdf $p(\hat m)$ for $\hat m$ and 
of the effective potential $\varOmega_N(\hat m,\beta)$
are straightforward
\begin{equation}\label{eq:Omega}
\begin{split}
p(\hat m) = \ee^{N \varOmega_N(\hat m, \beta)}= &\frac{1}{Z} \int\limits_{-\infty}^\infty \prod_{i=1}^N \dd\phi_i 
\sum_{\{\sigma_{\boldsymbol x}\}} \ee^{-\beta E + M - \hat M}\\
&\times \delta\biggl(\hat m - m - \sum_i \phi^2_i/(2N)\biggr).
\end{split}
\end{equation}
Our use of demons is reminiscent of
Creutz's microcanonical algorithm~\cite{CREUTZ}, 
but we shall integrate the $\phi_i$ out
in order to define our \emph{tethered averages},
\begin{equation}
\langle O\rangle_{\hat m,\beta} = \frac{ \sum_{\{\sigma_{\boldsymbol x}\}} O(\hat m; \{\sigma_{\boldsymbol x}\})
\ \omega_N(\beta,\hat m; \{\sigma_{\boldsymbol x}\})}{\sum_{\{\sigma_{\boldsymbol x}\}} 
\ \omega_N(\beta,\hat m; \{\sigma_{\boldsymbol x}\})},
\end{equation}
where $O$ represents a generic observable and
\begin{equation}\label{eq:omega}
\omega_N(\beta,\hat m; \{\sigma_{\boldsymbol x}\}) = \ee^{-\beta E+M-\hat M} (\hat m-m)^{(N-2)/2} \theta(\hat m -m),
\end{equation}
($\theta$ is Heaviside's step function). Then we find that
\begin{equation}\label{eq:hath}
\hat h \equiv  - 1 + \frac{N/2-1}{\hat M - M}, \qquad 
\langle\hat h\rangle_{\hat m,\beta}= \frac{\partial \varOmega_N(\hat m,\beta)}{\partial \hat m}.
\end{equation}
Therefore, we can construct the effective potential by 
integrating $\langle \hat h\rangle_{\hat m,\beta}$ on $\hat m$.
Once we have $\varOmega_N(\hat m,\beta)$ 
we can compute canonical averages for any given value
of the external magnetic field $h$ with the formula
\begin{equation}\label{eq:canonical-average}
\langle O \rangle_\beta(h) = \frac{\int \dd \hat m\ \ee^{N[\varOmega_N(\hat m,\beta)+h \hat m]} \langle O\rangle_{\hat m,\beta}}
{\int \dd \hat m\ \ee^{N[\varOmega_N(\hat m,\beta)+h \hat m]}}\ .
\end{equation}

\begin{figure}[b]
\includegraphics[height=\columnwidth,angle=270]{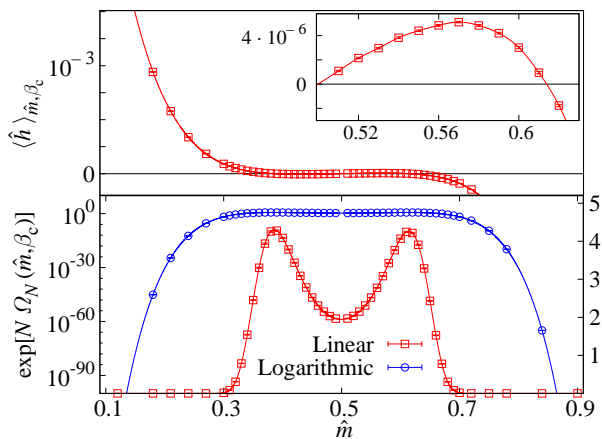}
\caption{(color online).
 Computation of the effective potential $\varOmega_N(\hat m ,\beta_\text{c})$, eq.~\eqref{eq:Omega},
for an $L=128$ Ising $3D$ lattice.
\textbf{Top:}
We simulate each point independently  to measure
$\langle \hat h\rangle_{\hat m,\beta_\text{c}}$, eq.~\eqref{eq:hath},
(the lines are cubic 
splines and the errors are much smaller than the points).
\textbf{Bottom:} This curve is then integrated to yield $p(\hat m)$, 
which we show in a linear ($\square$) and in a logarithmic
($\bigcirc$) scale. \textbf{Inset:} $\langle \hat h\rangle_{\hat m,\beta_\text{c}}$
for $\hat m \simeq m +1/2 >1/2$.}
\label{fig:tethered}
\end{figure}

The TMC simulation algorithm consists of four steps:
(1) Select an appropriate sampling $\hat m_i$, $i=1,\ldots,N_{\hat m}$ for $\hat m$, 
keeping in mind that $\hat m \simeq m + 1/2$. (2)
Run independent simulations for each $\hat m_i$, measuring
the tethered averages $\langle O\rangle_{\hat m,\beta}$.
(3) Integrate (a smooth interpolation of) $\langle \hat h\rangle_{\hat m,\beta}$ to 
obtain $\varOmega_N(\hat m,\beta)$. (4) Use equation~\eqref{eq:canonical-average} to 
recover the canonical averages.
Figure~\ref{fig:tethered} illustrates this process for the
$3D$ Ising model.
Let us remark that this is the algorithm for reproducing
canonical averages. However, see below, one can 
also obtain physically relevant results directly from
the tethered averages.

In \cite{TMC} we implemented this algorithm,
using Metropolis dynamics for step (2). Surprisingly enough,
magnetic observables such as $\hat h$ or $m$ presented no critical
slowing down (other quantities such as the energy presented
the $z \approx 2$ behavior typical of a local update~\cite{BRAY-TAMAYO-KLEIN}). 

Our cluster algorithm, a tethered version of Swendsen-Wang, 
is best explained using the bond-occupation variables $n_{\boldsymbol x \boldsymbol y}$ $(=0,1)$ 
and the conditional probability  distributions of~\cite{EDWARDS-SOKAL}. The lattice
bond joining neighboring sites $\boldsymbol x$ and $\boldsymbol y$ is \emph{occupied}
if $n_{\boldsymbol x \boldsymbol y}=1$. The occupied bonds partition the lattice
in connected clusters of size $N_i$, $i=1,\ldots,N_\text{c}$. Plugging in
\eqref{eq:omega} the identity
\begin{equation}\label{eq:bonds}
\ee^{\beta (\sigma_{\boldsymbol x} \sigma_{\boldsymbol y} -1)}
= \sum_{n_{\boldsymbol x\boldsymbol y} = 0,1}
[ (1-p) \delta_{n_{\boldsymbol x \boldsymbol y,0}} + p  \delta_{\sigma_{\boldsymbol x}, \sigma_{\boldsymbol y}}
\delta_{n_{\boldsymbol x \boldsymbol y,1}}], 
\end{equation}
where $p= 1- \ee^{-2\beta}$,
we immediately read the conditional probabilities of the spins
given the bonds and vice versa: (a) As in a canonical case,
 given the $\{\sigma_{\boldsymbol x}\}$, bonds are independent and 
$n_{\boldsymbol x\boldsymbol y}$ is 1 with probability
$p\delta_{\sigma_{\boldsymbol x},\sigma_{\boldsymbol y}}$. (b)
Given the $n_{\boldsymbol x\boldsymbol y}$ the $N_i$ spins within
cluster $i$ are equal to $S_i=\pm 1$.
The probability of the  $2^{N_\text{c}}$ $\{S_i\}$ configurations 
depends on $M = \sum_{i=1}^{N_\text{c}} S_iN_i$ through eq.~\eqref{eq:omega}:
\begin{equation}\label{eq:prob-clusters}
p(\{S_i\})\propto \ee^{M-\hat M} (\hat M-M)^{(N-2)/2}\, \theta(\hat M -M).
\end{equation}

Our cluster update will consist, then, in a cluster tracing using
conditioned probability (a) and in a cluster flipping using (b). For
this last step we perform a dynamical Monte Carlo, taking the $S_i$ at
$t=0$ from the initial spin configuration.  At each $t$ we select
$N'_\text{c}\ll N_\text{c}$ clusters ($N'_\text{c} \approx 5$ works
just fine).  We randomly pick lattice sites, selecting the cluster to
which they belong, until we find $N'_\text{c}$ different clusters.  We
then use~\eqref{eq:prob-clusters} to perform a heat bath among the
$2^{N'_\text{c}}$ configurations with the remaining
$N_\text{c}-N'_\text{c}$ clusters fixed \footnote{Let $k$ be one of
  the $2^{N'_\text{c}}$ assignments, and $f_k$ be its
  weight~\eqref{eq:prob-clusters}.  Form the vector $A_k=\sum_{r=1}^k
  f_r/ \sum_{r=1}^{2^{N'_\text{c}}} f_r$, with $A_{-1}=0$ . Extract a
  uniform random number $0\leq R<1$. For our chosen $k$, $ A_{k-1}\leq
  R < A_k$.}.  We take $N_\text{rep}$ such steps.
\begin{figure}[b]
\includegraphics[height=\linewidth,angle=270]{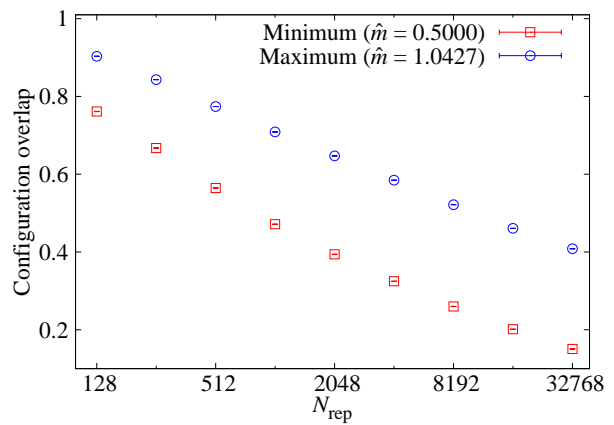}
\caption{(color online).
Spin overlap, eq.~\eqref{eq:overlap}, in a $2D$ critical $L=512$ Ising model
at the central minimum ($\square$) and right maximum ($\bigcirc$) of the effective
potential~\eqref{eq:Omega}.} 
\label{fig:overlap}
\end{figure}
\begin{table*}[t]
\begin{tabular*}{\textwidth}{@{\extracolsep{\fill}}lllcllll}
\multicolumn{3}{c}{$D=2$} &\quad &
\multicolumn{4}{c}{$D=3$} \\
\cline{1-3}\cline{5-8}
$L$ & Cluster & Met. + Cluster & &  $L$ & Cluster & Met. + Cluster & Swendsen-Wang \\
\cline{1-3}\cline{5-8}
 16  & 2.310(14) & 0.775(3)&           &  16      & 2.135(13)   & 0.782(3)     & 5.459(3)  \\ 
 24  & 2.440(26) & 0.920(4)&           &  32      & 2.80(3)     & 1.134(5)     & 7.963(9)  \\    
 32  & 2.758(20) & 1.055(5)&           &  48      & 3.467(28)   & 1.427(8)     & 9.831(9)  \\    
 64  & 3.347(22) & 1.417(7)&           &  64      & 3.88(3)     & 1.700(10)    & 11.337(12)  \\  
 128 & 4.11(5)   &1.861(12)&           &  96      & 4.79(5)     & 2.152(14)    & 13.90(3)    \\        
 256 & 4.87(4)   &2.391(16)&           &  128     & 5.46(6)     & 2.566(17)    & 15.90(5)  \\
 512 & 5.79(8)   &3.040(24)&           &  192     & 6.54(11)    & 3.32(4)     & 19.10(9)  \\        
 1024& 6.78(8)   &3.70(4)  &           &  256     & 7.51(13)    & 3.85(5)      & 21.83(10)  \\
\cline{1-3}\cline{5-8}                   
$z_E$  & 0.241(7) &        &           & $z_E$     & 0.472(8)& 0.591(4)  & 0.460(5)  \\
$\chi^2$/d.o.f. & 0.36/2   &           &           & $\chi^2$/d.o.f.     & 5.85/5  & 4.61/5       &   \\           
\cline{1-3}\cline{5-8}
\end{tabular*}  
\caption{Integrated autocorrelation times for the energy at $\hat m= 0.5$ and $\beta = \beta_c$
for the $D=2,3$ Ising model. We compare  the cluster 
and mixed versions of our TMC algorithm. We also include
the results of~\cite{OSSOLA-SOKAL} for canonical Swendsen-Wang. Our values for $z_E$ 
are fits to $\tau_E = AL^{z_E}$, where the smallest lattice in range
is $L_\text{min} = 128 $ (2$D$) and $L_\text{min} = 32$ (3$D$).
}\label{tab:tau}
\end{table*}

Taking $N_\text{rep} > 1$ steps is advantageous because
over a large number of repetitions many of the $S_i$ will eventually be
flipped, decorrelating the system. Furthermore,
it takes much longer to trace the clusters than to flip
them once, so $N_\text{rep}$ can be made
relatively large without noticeably increasing the simulation time.
In Figure~\ref{fig:overlap} we represent the overlap
\begin{equation}\label{eq:overlap}
o = \frac{\bigl\langle \sum_{\boldsymbol x} [\sigma_{\boldsymbol x}^{t=0} \sigma_{\boldsymbol x}^{t=N_\text{rep}}
 - \langle m\rangle_{\hat m,\beta}^2]\bigr\rangle_{\hat m,\beta}}{N(1-\langle m\rangle^2_{\hat m,\beta})},  
\end{equation}
which vanishes for completely uncorrelated configurations.
Clearly, the configuration can significantly evolve
for a fixed distribution of the bonds. A major 
error reduction (a factor $25$ in our largest lattices) is achieved
by measuring $\hat h$ at \emph{each} of the $N_\text{rep}$ steps.

Naturally, one must eventually refresh the bond configuration. A
simulation at fixed $\hat m$ then consists of $N_\text{MC}$ steps
where one traces the clusters and performs $N_\text{rep}$ iterations
of the random walk in the $S_i$ space. We have empirically
found that an $N_\text{rep}$ that equilibrates the cluster-tracing
and cluster-flipping times is close to optimal, and very easy to find
(one only has to scale $N_\text{rep}$ with $N$, as 
the tracing of clusters is an $\mathcal O(N \ln N)$ operation).
For $N=256^3$ spins, this results in $N_\text{rep} \approx 5\cdot10^5$.

The efficiency of a MC method is best assessed through the equilibrium
autocorrelation function~\cite{SOKAL} for an observable~$O$,
$C_{O}(t) \equiv \bigl\langle[O(0)-\langle O\rangle] [ O(t)-\langle O\rangle] \bigr\rangle$.
The slowness of the dynamics can be quantified with the integrated autocorrelation
time $\tau_{O}$. Defining $\rho_O(t)=C_O(t)/C_O(0)$, $\tau_O$ 
is just the time integral of $\rho_O(t)$ in $(0,\infty)$. We estimate
it with the self-consistent window method~\cite{SOKAL} for our
numerical estimate $\bar \rho_O$ of $\rho_O$, 
\begin{equation}\label{eq:tau}
\tau_O = \frac12 + \sum_{t=1}^{\varLambda} \bar\rho_O(t),\quad \varLambda = W \tau_O.
\end{equation}
We typically use $W=6$, but have checked that the results are consistent
for several $W$'s. Since $\tau_O\propto L^{z_O}$, one is interested in the
observable with largest $z_O$ ($E$ in our case,
as is typical for cluster methods).

We have used $\tau_E$ to assess our algorithm's performance
for the Ising model in two
(at $\beta=\beta_\text{c} = \ln(1+\sqrt2)/2$) and three dimensions
(at $\beta = 0.22165459 \approx \beta_\text{c}$~\cite{BETAC}).
For both $D$'s, we find that the $\tau_E$ are 
largest at $\hat m = 0.5$ (the central minimum
of $p(\hat m)$), so we report their values at that
point in Table~\ref{tab:tau}.
In $2D$ we obtain $z_E = 0.241(7)$, while in
$3D$ our dynamic exponent is $z_E = 0.472(8)$, 
compatible with the Swendsen-Wang value of $z_E = 0.460(5)$~\cite{OSSOLA-SOKAL}.
We also include our results with a slightly modified algorithm, where
we take two Metropolis steps each cluster step. This mixed algorithm
has significantly smaller $\tau'_E$ in both dimensions.
Paradoxically, in $3D$, $z'_E> z_E$ ($z'_E$ is unmeasurable in $2D$), which probably means
that larger $L$ would be needed to compute it properly.
Since for our lattice sizes the mixed 
algorithm fares better,
we shall use it hereafter.

As a proof of TMC's accuracy, we have reproduced some of the
critical Swendsen-Wang simulations of~\cite{OSSOLA-SOKAL}
for the $3D$ Ising model (Table~\ref{tab:canonical-averages}).
In accordance with our $\tau_E$ analysis,
the errors with both algorithms are comparable.
In $2D$, the new cluster algorithm
outperforms Metropolis.
For instance, for an $L=1024$ lattice, taking 10 times fewer steps than in~\cite{TMC}
and the same $\hat m$ grid, cluster errors are 8 times smaller for $E$, 6 times
smaller for $C$ and 10 times smaller for $\chi$ and $\xi$. 

\begin{table}[h]
\begin{tabular*}{\columnwidth}{@{\extracolsep{\fill}}lccccc}
\cline{2-6}
& MCS & $-\langle e\rangle_\beta$ & $C$ & $\chi$ & $\xi$ \\
\hline
SW   & $48\times10^6$ & 0.3309822(16) & 22.155(18) & 21193(13) & 82.20(3)\\
TMC  & $50\times10^6$   & 0.3309831(15) & 22.174(13) & 21202(13) & 82.20(6)\\
\hline 
\end{tabular*}
\caption{Comparison of canonical Swendsen-Wang 
(data from~\cite{OSSOLA-SOKAL}) with TMC for an $N=128^3$ lattice at $\beta_\text{c}$.
We take $10^6$ MC steps at each of the $50$ points 
of our $\hat m$ grid (Figure~\ref{fig:tethered}). This results in a 
similar number of MCS for both simulations.
($\xi$: second-moment correlation length~\cite{COOPER-xi,VICTORAMIT}).}
\label{tab:canonical-averages}
\end{table}
  
Yet TMC can do more than reproduce canonical averages. Let 
us compute the anomalous dimension~$\eta$. 
Finite-size scaling \cite{VICTORAMIT,TMC} tells us that
the right maximum of $p(\hat m;L)$ at $\beta_\text{c}$
(Figure~\ref{fig:tethered}) scales as 
\begin{equation}\label{eq:peak}
\hat m_\text{peak} -\tfrac 12 =  A L^{-(\eta+D-2)/2}  + \ldots,\quad (A=\text{const.})
\end{equation}
where the dots stand for scaling corrections.

Unlike in a canonical simulation, we
locate $\hat m_\text{peak}$ through $\langle\hat h\rangle_{\hat m_\text{peak},\beta} = 0$.
We simulate two very close values of $\hat m$
at either side of the peak, and use a linear interpolation. 
Our $3D$ results are in Table~\ref{tab:eta}.

\begin{table}[t]
\begin{tabular*}{\columnwidth}{@{\extracolsep{\fill}}lllll}
\hline
\multicolumn{1}{c}{$L|L_\text{min}$}& \multicolumn{1}{c}{MCS} &  \multicolumn{1}{c}{$\hat m_\text{peak} -\tfrac12$} &  \multicolumn{1}{c}{$\eta$} & \multicolumn{1}{c}{$\chi^2$/d.o.f.}\\       
\hline
16  & $1.0\times10^8$ & 0.33421(5)  & 0.03392(21)          & 42.6/6\\ 
32  & $1.0\times10^8$ & 0.23377(4)  & 0.03526(30)          & 8.79/5\\
48  & $1.0\times10^8$ & 0.18956(4)  & \textbf{0.0360}(5)   & \textbf{4.24/4}\\
64  & $1.0\times10^8$ & 0.16341(4)  & 0.0368\textbf{(7)}   & 1.02/3\\
96  & $1.0\times10^8$ & 0.13240(4)  & 0.0363(12)           & 0.78/2\\
128 & $1.0\times10^8$ & 0.114083(24)& 0.0373(19)           & 0.31/1\\
192 & $6.0\times10^7$ & 0.09246(4)  & \multicolumn{1}{c}{---} & \multicolumn{1}{c}{---} \\
256 & $8.2\times10^6$ & 0.07959(12) & \multicolumn{1}{c}{---} & \multicolumn{1}{c}{---} \\
\hline
\end{tabular*}
\caption{
Position of the peak of $p(\hat m;L)$
for the $D=3$
Ising model at $\beta_\text{c}$. 
In each row we report 
the result of a fit to \eqref{eq:peak}
(the first column is also the fit's $L_\text{min}$).} 
\label{tab:eta} 
\end{table}

We have found that 
eq.~\eqref{eq:peak}  yields remarkably good fits for
$L_\text{min}= 48$. Furthermore, increasing $L_\text{min}$
results in compatible values of $\eta$, with growing errors.
Our preferred estimate is  $\eta = 0.0360(7)$,
where we took the central value from $L_\text{min}=48$
and the error from $L_\text{min} = 64$ to account 
for systematic effects. This estimate 
compares favorably with the best Monte Carlo computation
known to us, $\eta = 0.0362(8)$~\cite{HASENBUSCH}
and is compatible with a high-temperature expansion value of 
$\eta =0.03639(15) $~\cite{ETA-HT} (however, 
both quoted values~\cite{HASENBUSCH,ETA-HT}
were computed with a perfect action,
not in the Ising model).

In summary, we have shown how models with conserved 
order parameter can be efficiently simulated with
a cluster method. We work in 
the tethered ensemble framework, which allows us to compute
the Helmholtz effective potential. The method 
is tested in the $D=2,3$ Ising model. For the computation
of canonical expectation values in large lattices, our cluster algorithm
is no less efficient than a canonical one in $3D$
(in $2D$ the dynamical exponent $z\approx0.24$ is larger
than that for the canonical algorithm, but still very small).
The tethered ensemble permits a very efficient computation of quantities
such as the maxima of the effective potential, which would be extremely 
costly to reproduce in a canonical setting. 
Our estimate for the anomalous dimension of the $3D$
Ising model compares favorably with all previous Monte Carlo
computations known to us.
We plan to further develop this algorithm to study disordered
systems~\cite{DESORDENADOS} and the condensation transition~\cite{CONDENSACION}.

We thank L.A. Fernandez for discussions. We were partly 
supported by MEC (Spain) through contract FIS2006-08533.
DY is an FPU fellow (Spain). The bulk of our
computations were carried out at BIFI.

\end{document}